\documentclass[final,3p,12pt]{elsarticle}

\journal{Computer Physics Communications}

\usepackage{amssymb}
\usepackage{amsmath}
\usepackage{amsfonts}
\biboptions{sort&compress}

\usepackage{hyperref}
\usepackage{titlesec}
\usepackage{xcolor}

\titleformat*{\section}{\Large\bfseries}
\titleformat*{\subsection}{\large\bfseries}
\titlespacing*{\section}{0pt}{5.0ex}{4.0ex}
\titlespacing*{\subsection}{0pt}{3.0ex}{2.0ex}

\hypersetup{
    colorlinks,
    linkcolor={red!50!black},
    citecolor={blue!50!black},
    urlcolor={blue!80!black}
}

\newcommand{\somespace}{\vspace{0.2cm}}

\newcommand{\D}{\mathrm{d}}
\renewcommand{\O}{\mathcal{O}}
\renewcommand{\j}{\varphi}
\newcommand{\bs}[1]{\boldsymbol{#1}}
\newcommand{\<}{\langle}
\renewcommand{\>}{\rangle}
\newcommand{\lla}{\langle \! \langle}
\newcommand{\rra}{\rangle \! \rangle}

\newcommand{\lwick}{:\!}
\newcommand{\rwick}{\!:}

\DeclareMathOperator{\K}{\mathrm{K}}

\DeclareMathOperator{\re}{Re}
\DeclareMathOperator{\Li}{Li}
\DeclareMathOperator{\NL}{NL}

\begin{document}

\begin{frontmatter}

\title{TripleK: A Mathematica package for evaluating triple-K integrals and conformal correlation functions}

\author[a]{Adam Bzowski\corref{author}}
\cortext[author] {\textit{E-mail address:} adam.bzowski@physics.uu.se}
\address[a]{Department of Physics and Astronomy, Uppsala University, 751 08 Uppsala, Sweden}

\begin{abstract}

\noindent I present a Mathematica package designed for manipulations and evaluations of triple-$K$ integrals and conformal correlation functions in momentum space. Additionally, the program provides tools for evaluation of a large class of 2- and 3-point massless multi-loop Feynman integrals with generalized propagators. The package is accompanied by five Mathematica notebooks containing detailed calculations of numerous conformal 3-point functions in momentum space.

\end{abstract}


\end{frontmatter}

\noindent {\bf PROGRAM SUMMARY}

\somespace

\begin{small}
\noindent {\em Program Title:} TripleK  \somespace \\
\noindent {\em Version:} 1.0 \somespace \\
\noindent {\em Program obtainable from:} \href{https://triplek.hepforge.org/}{https://triplek.hepforge.org/} \somespace \\
\noindent {\em License:} GNU General Public License v3.0 \somespace \\
\noindent {\em Programming language:} Wolfram Language \cite{Mathematica} (Mathematica 10.0 or higher) \somespace \\
\noindent {\em Computers:} Computers running Mathematica \somespace \\
\noindent {\em Distribution format:} tar.gz \somespace \\
\noindent {\em Other package needed:} none \somespace \\
\noindent {\em External file required:} none \somespace \\
\noindent {\em No. of bytes in distributed program, including test data, etc.:} 3.2 MB \somespace \\
\noindent {\em No. of lines in distributed program, including test data, etc.:} 82,700 \somespace \\
\noindent {\em Supplementary material:} The package includes five Mathematica notebooks containing bulk of the results regarding the structure of conformal 3-point functions. \somespace \\
\noindent {\em Nature of problem(approx. 50-250 words):}
Triple-$K$ integrals were introduced in \cite{Bzowski:2013sza} as a convenient tool for the analysis of conformal 3-point functions in momentum space. All 3-point functions of scalar operators, conserved currents and stress tensor can be expressed in terms of triple-$K$ integrals. Furthermore, a large class of 2- and 3-point massless multi-loop Feynman integrals with generalized propagators is expressible in terms of triple-$K$ integrals as well. Since the expressions are usually long and unwieldy, an automated tool is essential for efficient manipulations. \somespace \\
\noindent {\em Solution method(approx. 50-250 words):} In \cite{Bzowski:2015yxv} an effective reduction algorithm was provided for expressing a large class of triple-$K$ integrals in terms of master integrals. The presented package implements this reduction scheme. As far as the multi-loop Feynman integrals are concerned, the conversion to multiple-$K$ integrals proceeds by means of Schwinger parameterization. \somespace \\
\noindent {\em Additional comments including Restrictions and Unusual features (approx. 50-250 words):} Despite extensive testing, this package is a one man job, therefore bugs are unavoidable. Please, report all issues at \textit{adam.bzowski@physics.uu.se} or \textit{abzowski@gmail.com}. \somespace \\
\noindent {\em Restrictions on complexity of the problem:} Limited by memory and time available. \somespace \\
\noindent {\em Typical running time:} Strongly depending on the problem. Evaluation of all attached notebooks: ca. 35 minutes. \somespace \\

\noindent \cite{Mathematica} Wolfram Research Inc., \href{https://www.wolfram.com/mathematica}{Mathematica}, Version 11.2, 12.0, Champaign, IL, 2020
\\
\noindent \cite{Bzowski:2013sza} A.~Bzowski, P.~McFadden, K.~Skenderis, {Implications of conformal invariance in
  momentum space}, JHEP 03 (2014) 111.
\newblock \href {http://arxiv.org/abs/1304.7760} {\path{arXiv:1304.7760}},
  \href {https://doi.org/10.1007/JHEP03(2014)111}
  {\path{doi:10.1007/JHEP03(2014)111}} \\
\noindent \cite{Bzowski:2015yxv} A.~Bzowski, P.~McFadden, K.~Skenderis, {Evaluation of conformal integrals},
  JHEP 02 (2016) 068.
\newblock \href {http://arxiv.org/abs/1511.02357} {\path{arXiv:1511.02357}},
  \href {https://doi.org/10.1007/JHEP02(2016)068}
  {\path{doi:10.1007/JHEP02(2016)068}}  \\

\end{small}

\newpage

\tableofcontents

\newpage

\section{Introduction} \label{sec:intro}

Problem of analytical or numerical evaluation of Feynman diagrams has been at the heart of high energy research and has been tackled by numerous authors. A number of packages designed for analytic evaluation and manipulations of amplitudes and Feynman diagrams exists. In most cases, programs are delivered as Mathematica packages, most suitable for symbolic manipulations. The most popular programs include \verb|FeynCalc|, \cite{Mertig:1990an,Shtabovenko:2020gxv}, \verb|Package-X|, \cite{Patel:2015tea,Patel:2016fam}, \verb|LoopTools|, \cite{Hahn:1998yk}, \verb|HEPMath|, \cite{Wiebusch:2014qba}, \verb|FIRE|, \cite{Smirnov:2008iw,Smirnov:2019qkx}, \verb|LiteRed|, \cite{Lee:2012cn}, and more.

The main focus of the standard set-up is to consider Feynman diagrams in (close to) 4 spacetime dimensions and composed from a number of bosonic or fermionic massive propagators, with the usual $1/(p^2 + m^2)$ factor. However, due to new developments in momentum space conformal field theory, a similar but different problem has arisen. As shown recently, \cite{Bzowski:2019kwd}, all conformal (scalar) correlation functions in momentum space can be expressed as momentum integrals with massless, generalized propagators $1/p^{2 \nu}$, with $\nu$ not necessarily equal to one. Furthemore, applications in condensed matter physics, cosmology, or string theory, require the analysis to take place in a wide set of spacetime dimensions. While standard methods employed by the aforementioned programs can be used to some extent in the analysis of such problems, new methods aimed specifically at conformal correlators can be developed.

As far as conformal 2- and 3-point functions are concerned, in \cite{Bzowski:2013sza} a novel approach was proposed, by expressing the correlators in terms of \emph{triple-K} integrals,
\begin{align} \label{pre_tripleK}
I_{\alpha \{\beta_1 \beta_2 \beta_3\}}(p_1, p_2, p_3) = p_1^{\beta_1} p_2^{\beta_2} p_3^{\beta_3} \int_0^\infty \D x \, K_{\beta_1}(p_1 x) K_{\beta_2}(p_2 x) K_{\beta_3}(p_3 x),
\end{align}
where $\alpha$ and $\beta_1, \beta_2, \beta_3$ are parameters related to the dimensions of the operators involved and $p_1, p_2, p_3$ are magnitudes of momenta $\bs{p}_1, \bs{p}_2$ and $\bs{p}_3 = - \bs{p}_1 - \bs{p}_2$. Furthermore, $K_\nu(z)$ denotes the modified Bessel function of the third kind. In subsequent papers \cite{Bzowski:2015pba,Bzowski:2017poo,Bzowski:2018fql} a detailed analysis of 3-point functions involving scalar operators as well as conserved currents and stress tensors was presented. 

A large class of physically significant triple-$K$ integrals can be expressed analytically. To achieve it, a comprehensive algorithm was presented in \cite{Bzowski:2015yxv}. In this paper I introduce a Mathematica package, which implements this algorithm. From the point of view of Feynman diagramatics the package provides tools for evaluation of 2- and 3-point massless multi-loop Feynman diagrams with generalized propagators. In addition, the package includes a number of notebooks containing results constituting bulk of the material published in \cite{Bzowski:2013sza,Bzowski:2015pba,Bzowski:2017poo,Bzowski:2018fql}.

\section{Physical significance}

Triple-$K$ integrals \eqref{pre_tripleK} were introduced in \cite{Bzowski:2013sza} as a convenient tool for the analysis of conformal 3-point functions in momentum space. Their significance comes from the fact that they provide natural way of expressing solutions to conformal Ward identities. This includes any 2- and 3-point functions of conformal operators of arbitrary spin such as conserved currents or stress tensor.

Furthermore, a large class of triple-$K$ integrals can be analytically expressed in terms of almost elementary functions. In particular all 3-point functions of operators of integral conformal dimensions in odd-dimensional spacetimes can be evaluated explicitly in terms of rational functions of momenta magnitudes only. In case of 3-point functions of operators of integral dimensions in even-dimensional spactimes, triple-$K$ integrals provide a reduction scheme which leads to analytic expressions containing single special function: dilogarithm.

Finally, triple-$K$ integrals can be used for explicit evaluation of massless 3-point Feynman diagrams. All such momentum loop integrals can be expressed in terms of triple-$K$ integrals, which then can be turned into explicit expressions. This provides new analytic expressions for a large class of Feynman diagrams.

\subsection{Conformal invariance}

On the level of correlation functions conformal invariance manifests itself through conformal Ward identities. In addition to known consequences of Poincar\'{e} invariance, conformal invariance imposes further constrains through dilatation Ward identity and a set of special conformal Ward identities.

Consider a general $n$-point function in momentum space of arbitrary operators $\O_1, \ldots, \O_n$ of conformal dimensions $\Delta_1, \ldots, \Delta_n$. We work in $d$ Euclidean spacetime dimensions and assume $d > 2$. Let us consider the $n$-point function in momentum space and introduce the double bracket notation \textit{via}
\begin{align}
\< \O_1(\bs{p}_1) \ldots \O_n(\bs{p}_n) \> = (2 \pi)^d \delta(\bs{p}_1 + \ldots + \bs{p}_n) \lla \O_1(\bs{p}_1) \ldots \O_n(\bs{p}_n) \rra,
\end{align}
The $n$-point function then depends on $n-1$ independent momenta, with $\bs{p}_n = - (\bs{p}_1 + \ldots + \bs{p}_{n-1})$. With notation in place, the dilatation Ward identity simply forces the $n$-point function to be a homogeneous function of dimension $\Delta_t - (n-1)d$, where $\Delta_t = \sum_{j=1}^n \Delta_j$. 

Special conformal Ward identities comprise of a set of second-order differential equations labeled by a single index, $\kappa$. Their exact form depends on the tensor structure of the operators involved and can be schematically written as
\begin{align} \label{CWIOp}
\left( \mathcal{K}^{\kappa} \bs{1} + \mathcal{\bs{T}}^{\kappa} \right) \lla \O_1(\bs{p}_1) \ldots \O_n(\bs{p}_n) \rra = 0,
\end{align}
where $\mathcal{K}^{\kappa}$ is a second-order differential operator independent of the tensor structure, while $\mathcal{\bs{T}}^{\kappa}$ is a first-order differential operator depending on the tensor structure of the operators. The CWI operator $\mathcal{K}^{\kappa}$ equals
\begin{align} \label{CWIOpScal}
\mathcal{K}^\kappa = \sum_{j=1}^{n-1} \left[ p_j^\kappa \frac{\partial}{\partial p_j^\alpha} \frac{\partial}{\partial p_{j \alpha}} - 2 p_j^\alpha \frac{\partial}{\partial p_j^\alpha} \frac{\partial}{\partial p_{j \kappa}}+2(\Delta_j-d)\frac{\partial}{\partial p_{j \kappa}} \right],
\end{align}
while the explicit form of $\mathcal{\bs{T}}^{\kappa}$ can be found in \cite{Bzowski:2013sza}.

By carrying out a suitable decomposition of the tensorial structure, one can rewrite conformal Ward identities as a set of scalar Ward identities. This has been carried out for 2- and 3-point functions of scalar operators, conserved currents and stress tensor in \cite{Bzowski:2013sza,Bzowski:2017poo}. The second-order differential operator featuring prominently in these expressions is
\begin{align} \label{KOp}
\K_j(\beta) = \frac{\partial^2}{\partial p_j^2} - \frac{2 \beta - 1}{p_j} \frac{\partial}{\partial p_j}.
\end{align}
Note that in this operator the derivatives are taken with respect to the momentum magnitudes, $p_j$, $j=1,2,3$, \textit{i.e.}, $p_j = | \bs{p}_j |$. Due to the Poincar\'{e} symmetry any 3-point function in momentum space can be expressed in terms of three kinematic parameters, which can be taken to be the three momenta magnitudes.

As an example, consider the 3-point function of three scalar operators of dimensions $\Delta_1, \Delta_2, \Delta_3$. One finds two independent equations expressing special conformal Ward identities, which can be collectively written as
\begin{align}
0 = \left[ \K_i(\beta_i) - \K_j(\beta_j) \right] \lla \O_1(\bs{p}_1) \O_2(\bs{p}_2) \O_3(\bs{p}_3) \rra,
\end{align}
where $\beta_j = \Delta_j - d/2$ and $i,j=1,2,3$. Their solution in terms of the triple-$K$ integral \eqref{pre_tripleK} is extremely simple. The 3-point function is uniquely determined up to a single multiplicative constant $C$ (OPE coefficient),
\begin{align} \label{3pt_scalar}
\lla \O_1(\bs{p}_1) \O_2(\bs{p}_2) \O_3(\bs{p}_3) \rra = C I_{\frac{d}{2} - 1 \{ \Delta_1 - \frac{d}{2}, \Delta_2 - \frac{d}{2}, \Delta_3 - \frac{d}{2}\} }(p_1, p_2, p_3).
\end{align}
The value of the $\alpha$-parameter is determined by the dilatation Ward identity.

When spinning operators are considered, the 3-point function must first be decomposed into a set of tensors multiplying scalar form factors. Such decompositions were worked out in \cite{Bzowski:2013sza,Bzowski:2017poo,Bzowski:2018fql} for 3-point functions containing scalar operators, conserved currents and stress tensor. When the special Ward identities are applied to the decomposition they produce a set of differential equations obeyed by the form factors. Those in turn split into second-order differential equations called primary Ward identities and first-order differential equations called secondary Ward identities. Primary Ward identities can be solved in terms of triple-$K$ integrals, while secondary Ward identities impose additional constraints on the set of integration constants.

Solutions to the primary Ward identities can be expressed in terms of triple-$K$ integrals as in equation \eqref{3pt_scalar}, with $\alpha$ and $\beta$-indices shifted by integers. For this reason it is convenient to follow notation of \cite{Bzowski:2013sza} and define reduced integrals,
\begin{align} \label{Jint}
J_{N, \{ k_1 k_2 k_3 \}} = I_{N + \frac{d}{2} - 1 \{ k_1 + \Delta_1 - \frac{d}{2}, k_2 + \Delta_2 - \frac{d}{2}, k_3 + \Delta_3 - \frac{d}{2}\} },
\end{align}
where the values of $\Delta_1, \Delta_2, \Delta_3$ are implicitly assumed to be that of the conformal dimensions of the operators involved.

\subsection{Loop integrals}

Motivated by conformal invariance, one can define multiple-$K$ integrals by integrating a product of modified Bessel functions of the third kind,
\begin{align} \label{multipleK}
I_{\alpha \{ \beta_1, \ldots \, \beta_n \}}(p_1, \ldots, p_n) = \int_0^\infty \D x \, x^\alpha \prod_{j=1}^n p_j^{\beta_j} K_{\beta_j} (p_j x).
\end{align}
While all such expressions are conformal in momentum space, for $n > 3$ they only represent very special correlation functions, \cite{Bzowski:2019kwd,Maglio:2019grh}. However, double- and triple-$K$ integrals represent all conformal 2- and 3-point functions.

From the point of view of Feynman diagramatics it turns out that every 1-loop 2- and 3-point momentum integral of the form
\begin{align} \label{23pt_int}
& \int \frac{\D^d \bs{k}}{(2 \pi)^d} \frac{k^{\mu_1} \ldots k^{\mu_m}}{k^{2 \delta_1} |\bs{k} - \bs{p}|^{2 \delta_2}}
, && \int \frac{\D^d \bs{k}}{(2 \pi)^d} \frac{k^{\mu_1} \ldots k^{\mu_m}}{k^{2 \delta_3} |\bs{k} - \bs{p}_1|^{2 \delta_2} |\bs{k} + \bs{p}_2|^{2 \delta_1}}
\end{align}
can be expressed in terms of double- and triple-$K$ integrals. For example,
\begin{align} \label{3pt_to_i}
\int \frac{\D^d \bs{k}}{(2 \pi)^d} \frac{1}{k^{2 \delta_3} |\bs{k} - \bs{p}_1|^{2 \delta_2} |\bs{k} + \bs{p}_2|^{2 \delta_1}} = \frac{2^{4-\frac{3d}{2}}}{\pi^{\frac{d}{2}}} \times \frac{I_{\frac{d}{2} - 1 \{ \frac{d}{2} + \delta_1 - \delta_t, \frac{d}{2} + \delta_2 - \delta_t, \frac{d}{2} + \delta_3 - \delta_t \}}}{\Gamma(d-\delta_t) \Gamma(\delta_1) \Gamma(\delta_2) \Gamma(\delta_3)},
\end{align}
where $\delta_t = \delta_1 + \delta_2 + \delta_3$, while general expressions with arbitrary numerators can be found in \cite{Bzowski:2013sza,Anninos:2019nib}. 

There are numerous advantages in using multiple-$K$ integrals in place of usual loop momentum integrals:
\begin{itemize}
\item All parameters are scalars.
\item Bose symmetries (permutations of momenta) are manifest.
\item The result is expressed in terms of a single integral rather than a $d$-dimensional integral, which is much more convenient for numerical analysis.
\item Various identities between momentum integrals can be traced back to identities between Bessel functions.
\item Renormalization properties are in 1-to-1 correspondence with singularities of multiple-$K$ integrals, which in turn are easy to analyze.
\item Analytic expressions can be obtained for a wide class of integrals.
\end{itemize}
It is the main objective of the presented package to implement the two last points.

\subsection{Divergences and regularization} \label{sec:divs}

Most of the physically interesting multiple-$K$ integrals exhibit singularities, which can be related to the singularities of correlation functions they represent. The position and structure of the singularities can be obtained by the analysis of properties of the Bessel functions. In particular, divergent terms can always be evaluated without evaluating the entire integral. 

Assuming all $p_j > 0$ and fixed, the multiple-$K$ integral \eqref{multipleK} converges if
\begin{align}
\re \alpha + 1 - \sum_{j=1}^n | \re \beta_j | > 0.
\end{align}
By using analytic continuation one can extend the definition of the multiple-$K$ integral to a larger set of parameters $\alpha$ and $\beta_j$. From now on we will consider only real $\alpha$ and $\beta$-parameters and we will refer to this analytic continuation as multiple-$K$ integral. In such case the function exhibits poles whenever there exists a list of signs $(\sigma_1 \ldots \sigma_n)$ with $\sigma_j = \pm 1$ such that
\begin{align} \label{defn}
n_{(\sigma_1 \ldots \sigma_n)} = - \frac{1}{2} \left[ \alpha + 1 + \sum_{j=1}^n \sigma_j | \beta_j | \right]
\end{align}
is a non-negative integer. If the condition holds for some choice of signs $(\sigma_1 \ldots \sigma_n)$ we say that the singularity of type $(\sigma_1 \ldots \sigma_n)$ appears.

The order of a pole equals to the number of different choices of the signs $(\sigma_1 \ldots \sigma_n)$ up to reshuffling. This means that the highest possible pole has order $n+1$. Furthermore, note that if both conditions $(-, \sigma_2 \ldots \sigma_n)$ and $(+, \sigma_2 \ldots \sigma_n)$ hold, then the corresponding value of $\beta_1$ must be integral.

In many physically relevant cases the multiple-$K$ integrals do become singular. In such cases regularization is required. Using generalized dimensional regularization we shift the parameters $\alpha$ and $\beta_j$ by amounts proportional to the regulator $\epsilon$ and series expand resulting expressions. Since the parameters depend on spacetime dimension $d$ as well as conformal dimensions $\Delta_j$ of the operators involved, this is the generalized dimensional regularization scheme.

\subsection{Evaluation of multiple-K integrals} \label{sec:mulK}

All double-$K$ integrals can be evaluated explicitly in terms of hypergeometric functions. However, in the context of physical correlation functions the conservation of momentum implies that the only relevant integrals satisfy $p_1 = p_2$ in \eqref{multipleK}. In such case one finds
\begin{align}
i_{\alpha \{ \beta_1 \beta_2 \}}(p, p) = \frac{2^{\alpha - 2} p^{\beta_1 + \beta_2 - \alpha - 1}}{\Gamma(\alpha + 1)} \prod_{\sigma_1, \sigma_2 = \pm 1} \Gamma \left( \frac{1}{2} (\sigma_1 \beta_1 + \sigma_2 \beta_2 + \alpha + 1) \right).
\end{align}

Other multiple-$K$ integrals do not admit analytic expressions in a generic case. In principle, they can be expressed in terms of generalized hypergeometric functions: Appell $F_4$ function in case of triple-$K$ integral, Lauricella functions in case of quadruple-$K$, \cite{Maglio:2019grh}, and so on. These expressions are not convenient neither for numerical nor analytical manipulations. In some cases, however, simplifications occur. For example as far as triple-$K$ integrals are considered, the following cases can be expressed in terms of more or less elementary functions:
\begin{itemize}
\item All integrals with half-integral $\beta_j$ parameters are expressible in terms of elementary functions and Euler gamma function.
\item All integrals with two out of three half-integral $\beta_j$ parameters can be expressed in terms of the hypergeometric ${}_2 F_1$ function.
\item All integrals of the form $I_{\nu+1 \{\nu\nu\nu\}}$ and $I_{\nu-1 \{\nu\nu\nu\}}$ are expressible in terms of Legendre (hypergeometric) functions.
\end{itemize}
In particular, the case of half-integral $\beta_j$ parameters arises in the analysis of 3-point functions of operators of integral dimensions $\Delta_j$ in odd-dimensional spacetimes.

Most importantly, a large class of triple-$K$ integrals with integral $\beta_j$ parameters can be expressed in terms of elementary functions and dilogarithm, $\Li_2$. To be specific, the following conditions must be satisfied: 
\begin{enumerate}
\item[i)] all $\alpha$ and $\beta_1,\beta_2,\beta_3$ are integral, and 
\item[ii)] all $n_{(++-)}, n_{(+-+)}, n_{(-++)} < 0$, where $n_{(\sigma_1 \sigma_2 \sigma_3)}$ is defined in \eqref{defn}. 
\end{enumerate}
All such integrals can be expressed in terms of a single master integral, $I_{0 \{111\}}$ and the appropriate reduction scheme has been introduced in \cite{Bzowski:2015yxv}. The resulting expressions depend on two functions,
\begin{align} \label{NL}
\NL & = \frac{\pi^2}{6} - 2 \log \frac{p_1}{p_3} \log \frac{p_2}{p_3} + \log X \log Y - \Li_2 X - \Li_2 Y, \\
\lambda & = - (p_1 + p_2 - p_3) (p_1 - p_2 + p_3) (-p_1 + p_2 + p_3) (p_1 + p_2 + p_3), \label{lambda}
\end{align}
where
\begin{align}
& X = \frac{- p_1^2 + p_2^2 + p_3^2 - \sqrt{\lambda}}{2 p_3^2}, && Y = \frac{- p_2^2 + p_1^2 + p_3^2 - \sqrt{\lambda}}{2 p_3^2}.
\end{align}
Physically, such cases arise from the analysis of 3-point functions of operators of integral dimensions in even-dimensional spacetimes.

As an example, the 3-point function of the operator $\j^2$ in the theory of free massless scalar $\j$ in $d=4$ dimensions is proportional to $I_{1\{000\}}$, which in turn equals $\NL/(2 \sqrt{\lambda})$. 

\section{The package}

The most recent version of the package can be downloaded from the hepforge repository at \href{https://triplek.hepforge.org/}{https://triplek.hepforge.org/}.

\subsection{Files}

The package consists of 2 Wolfram Language (\textit{i.e.}, Mathematica) files: \verb|TripleK.wl| and \verb|Konformal.wl| as well as 5 Mathematica notebooks: \verb|BasicExamples.nb|, \verb|DeriveCWIs.nb|, \verb|SolveCWIs.nb|, \verb|CheckCWIs.nb|, and \verb|FreeTheory.nb|. All files were evaluated and checked using Mathematica versions 11.2 and 12.0, \cite{Mathematica}.

\begin{itemize}
\item \verb|TripleK.wl| contains the heart of the package. This file contains procedures for manipulations and evaluations of triple-$K$ integrals.

\item \verb|Konformal.wl| is a repository containing results regarding the structure of conformal 3-point functions of scalar operators, conserved currents and stress tensor. It gathers results published in the sequence of papers \cite{Bzowski:2013sza,Bzowski:2015pba,Bzowski:2017poo,Bzowski:2018fql}.
\end{itemize}

Remaining Mathematica notebooks contain examples and evaluations of various problems as well as checks on the results stored in \verb|Konformal.wl|.

\begin{itemize}
\item \verb|BasicExamples.nb| contains examples regarding the use of the package \verb|TripleK.wl|. It contains 5 sections presenting the package. First two sections, \textit{Evaluation of triple-K integrals} and \textit{Momentum-space integrals} provide the detailed description of the package's functionality. Then 3 complete examples follow. 

Section \textit{Scalar integrals} contains calculations of renormalization properties of two conformal scalar 3-point functions of operators: \textit{i}) $\Delta_1 = \Delta_2 = \Delta_3 = 3$ in $d=3$ dimensions, and \textit{ii}) $\Delta_1 = 4, \Delta_2 = \Delta_3 = 3$ in $d=4$ dimensions. The results are exemplified by carrying our calculations of 3-point functions $\< \j^6 \j^6 \j^6 \>$ and $\< \j^4 \j^3 \j^3 \>$ in a free massless scalar theory. These results were used for deriving examples 8 and 9 as well as Appendix C in \cite{Bzowski:2015pba}.

In section $\< T^{\mu\nu} \O \O \>$ the 3-point function $\< T^{\mu\nu} \j^2 \j^2 \>$ is evaluated in the theory of free massless scalar field. $\j$ denotes the scalar field while $T^{\mu\nu}$ is the stress tensor. The calculations are carried out in $d=3$ and $d=4$ spacetime dimensions.

Finally, in section \textit{Chiral anomaly} the 3-point function $\< j^{\mu} j^{\nu} j^{\rho} \>$ is evaluated. We consider the theory of a single free Weyl fermion in $d=4$ and $j^\mu$ denotes the chiral current, $j^\mu = \bar{\psi}_{\dot{\alpha}} \bar{\sigma}^{\mu \dot{\alpha} \alpha} \psi_{\alpha}$. Only after the 3-point function is calculated we apply the external momentum $p_{1 \mu}$ (\textit{i.e.}, calculate the divergence) and recover the well-known ABJ anomaly.

\item \verb|DeriveCWIs.nb| contains a derivation of both primary and secondary conformal Ward identities as reported in \cite{Bzowski:2013sza}. This is done by the application of the full conformal Ward identity in momentum space to the general decomposition of various 3-point functions. Ten sections cover all 3-point functions of scalar operators, conserved currents and stress tensor. The first section, \textit{Example:TJJ}, provides a more detailed description of the procedure when applied to the 3-point function of stress tensor and two conserved currents.

\item \verb|SolveCWIs.nb| contains solutions to primary and secondary CWIs. These solutions were reported in the series of papers, \cite{Bzowski:2013sza,Bzowski:2015pba,Bzowski:2017poo,Bzowski:2018fql}. We confirm the results by substituting them back to conformal Ward identities. The analysis of correlators involving stress tensor and conserved currents includes the issue of regulating spurious singularities, as discussed in detail in \cite{Bzowski:2017poo}. The analysis of correlation functions involving scalar operators is carried out in a generic, singularity-free case only.

\item \verb|CheckCWI.nb| contains complete solutions to primary and secondary CWIs in $d=3$ and $d=4$ spacetime dimensions. In correlation functions involving scalar operators, we consider operators of dimensions $\Delta = 2, 4$ in $d=4$ and $\Delta = 1, 3$ in $d=3$. Furthermore, in all the cases a general structure of possible semilocal terms is derived. All results were published in \cite{Bzowski:2017poo,Bzowski:2018fql}.

\item \verb|FreeTheory.nb| contains evaluations of all relevant 2- and 3-point functions in theories of free massless scalars and fermions in $d=3$ and $d=4$ spacetime dimensions. Quantities such as central charges or Euler anomaly coefficients are calculated. All results were published in \cite{Bzowski:2017poo,Bzowski:2018fql}.
\end{itemize}

\subsection{Installation}

In the current version the package does not contain a dedicated installer. One can access the package files, \verb|TripleK.wl| and \verb|Konformal.wl| by Mathematica's \verb|Get| command:

\includegraphics*[trim=2.0cm 24.0cm 0cm 2.3cm]{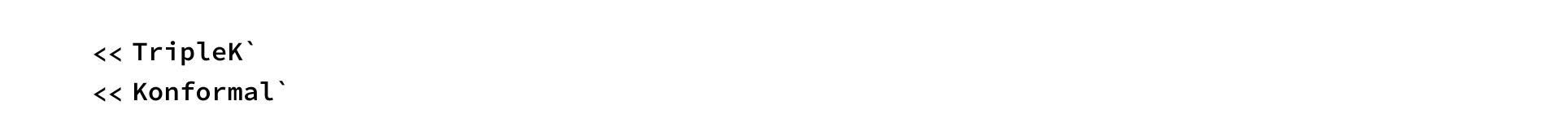}

\noindent Alternatively, one can install the packages in the Mathematica's folder structure using \verb|File -> Install| option from the Mathematica's menu.

\section{TripleK.wl}

\subsection{Basic manipulations}

Symbols 

\includegraphics*[trim=2.0cm 24.5cm 0cm 2.3cm]{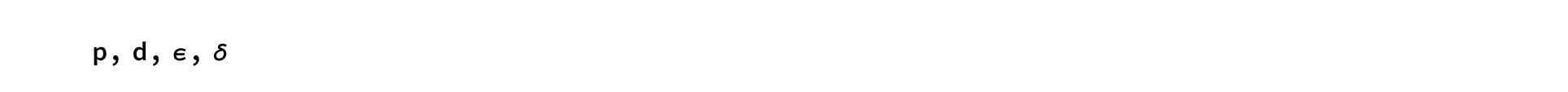}

\noindent are protected symbols introduced by the package. $\verb|p|$ is used for external momenta of triple-$K$ integrals. For $j=1,2,3$, \verb|p[j]| denotes the magnitude of the $j$-th momentum, $p_j = | \bs{p}_j |$, while \verb|p[j][|$\mu$\verb|]| denotes the actual vector, $p_j^\mu$. Only $\bs{p}_1$ and $\bs{p}_2$ are treated as independent momenta, while $\bs{p}_3 = - \bs{p}_1 - \bs{p}_2$. Euclidean metric $\delta^{\mu\nu}$ is denoted by $\delta$\verb|[|$\mu$\verb|,|$\nu$\verb|]|. Finally, \verb|d| denotes spacetime dimension.

The package provides basic functions for index manipulations. Indices can be contracted by calling:

\includegraphics*[trim=2.0cm 24.0cm 0cm 2.3cm]{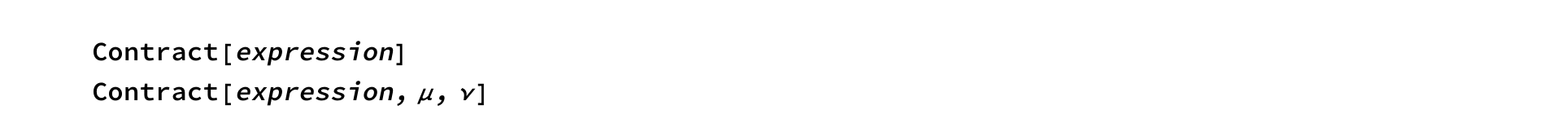}

\noindent In its first version all repeated indices in the expression will be contracted, provided they are located on recognized vectors. By default the only recognized vectors are \verb|p[j]| or those appearing under loop integrals. More vectors can be declared by adding the option \verb|Vectors -> v|, where \verb|v| is a single symbol or a list of symbols. 

In its second version \verb|Contract| contracts indices $\mu$ and $\nu$ in the expression. By default the contraction will take place over any symbols.

\verb|Contract| also admits a number of options. The most important option is \verb|Dimension|, which specifies the number of spacetime dimensions the contraction takes place in. If this option is not specified, symbol \verb|d| is used. For more information on \verb|Contract| use \verb|?Contract|. Various options to \verb|Contract| and more versions of \verb|Contract| are described in the attached Mathematica notebook \verb|BasicExpamples.nb|.

In order to differentiate a given expression with respect to the vector $k^\mu$ use

\includegraphics*[trim=2.0cm 24.5cm 0cm 2.3cm]{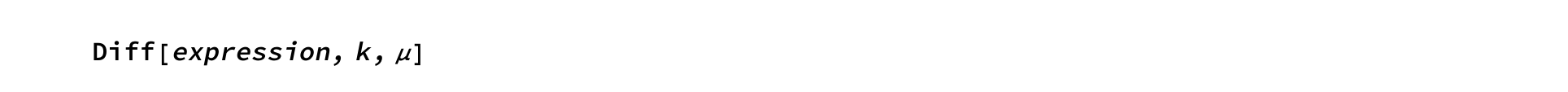}

\noindent If the vector $\bs{k}$ is equal to $\bs{p}_1$ or $\bs{p}_2$, every $\bs{p}_3$ in the expression is assumed to satisfy $\bs{p}_3 = - \bs{p}_1 - \bs{p}_2$ and the derivatives are taken accordingly.

\subsection{Evaluations of multiple-K integrals}

Double- and triple-\textit{K} integrals \eqref{multipleK} are represented as:

\includegraphics*[trim=2.0cm 24.0cm 0cm 2.3cm]{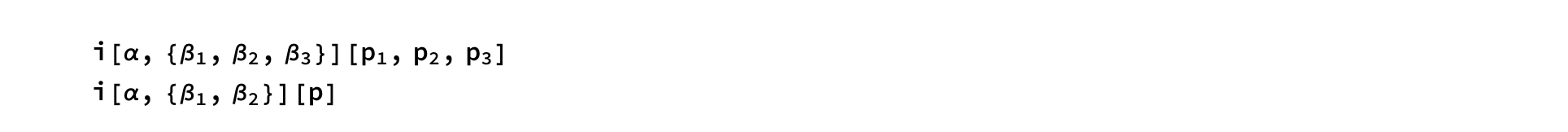}

\noindent In triple-\textit{K} integrals (but \textit{not} double-\textit{K}) momenta can be omitted. The standard momenta magnitudes \(p_1, p_2, p_3\)
are then assumed as arguments. In double-$K$ integrals only a single momentum magnitude appears as the argument. It is understood that $p_1 = p_2 = p$ as explained in section \ref{sec:mulK}. Parameters in the multiple-$K$ integrals can also be inputted as subscripts.

To evaluate all multiple-\textit{K} integrals in the given expression explicitly, use

\includegraphics*[trim=2.0cm 24.5cm 0cm 2.3cm]{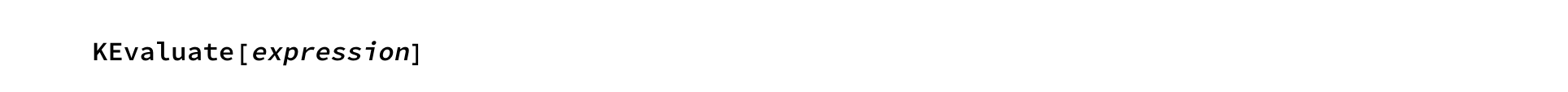}

\noindent This will replace all known triple-$K$ integrals by explicit expressions. Not all triple-$K$ integrals can be reduced to analytic expressions. The set of conditions which leads to analytic expressions is specified in section \ref{sec:mulK}. To check if a given triple-$K$ integral has an analytic representation available use

\includegraphics*[trim=2.0cm 24.5cm 0cm 2.8cm]{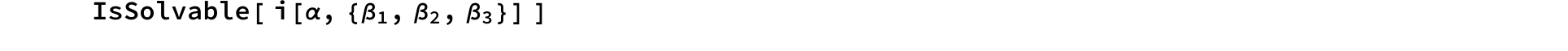}

\noindent which returns \verb|True| is the triple-$K$ integral can be reduced by \verb|KEvaluate| and \verb|False| otherwise.

Many interesting tripe-$K$ integrals diverge for a given set of $\alpha$ and $\beta$ parameters. In such case \verb|KEvaluate| produces a regulated expression, where $\epsilon$ is a protected symbol denoting the regulator. If the regulator is used explicitly in the expression the integral evaluates to a power series:

\includegraphics*[trim=2.0cm 23.5cm 0cm 2.2cm]{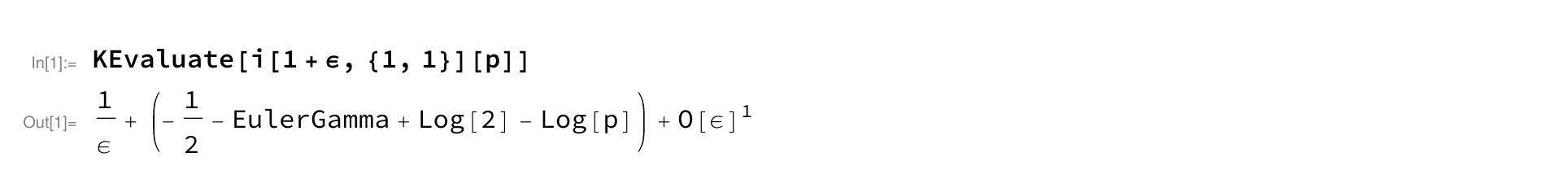}

\noindent If the integral diverges, but no regulator is specified, the default regularization 
\begin{align} \label{defreg}
\alpha \mapsto \alpha +u \epsilon, \qquad\qquad \beta _j \mapsto \beta _j+v \epsilon
\end{align}
is used, where \textit{u} and \textit{v} are arbitrary parameters:

\includegraphics*[trim=2.0cm 23.2cm 0cm 2.2cm]{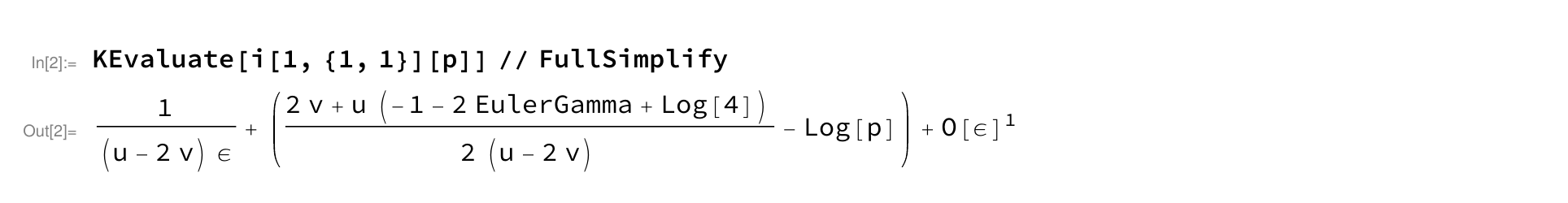}

The resulting expression can depend on two functions: \verb|NL| and $\lambda$ defined in \eqref{NL} and \eqref{lambda}. By default those functions are not expanded explicitly,

\includegraphics*[trim=2.0cm 23.2cm 0cm 2.2cm]{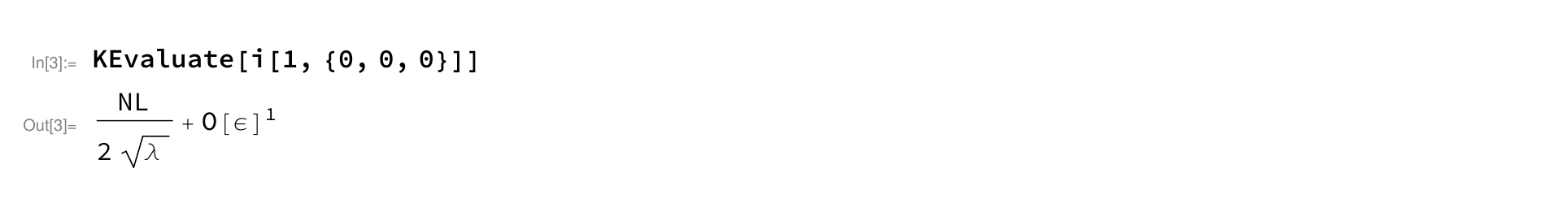}

\noindent To fully expand the two functions use \verb|KFullExpand|. We will describe this function in more detail in section \ref{sec:simp}.

\subsection{Divergences in multiple-K integrals}

The parts of triple-$K$ integrals that are divergent as $\epsilon$ approaches zero can be obtained without the evaluation of the entire integral. This is done by using

\includegraphics*[trim=2.0cm 24.5cm 0cm 2.3cm]{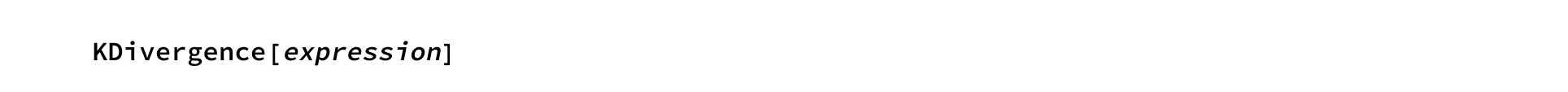}

\noindent As in case of \verb|KEvaluate|, the default regularization \eqref{defreg} is used if no explicit regularization is specified. By default \verb|KDivergence| evaluates all divergences together with scheme-dependent parts of the triple-$K$ integrals. Those are the terms that depend on the regularization scheme and can be used to change between various schemes.

The order of expansion can be adjusted by the option \verb|ExpansionOrder|. For example
\includegraphics*[trim=1.0cm 23.5cm 0cm 2.2cm]{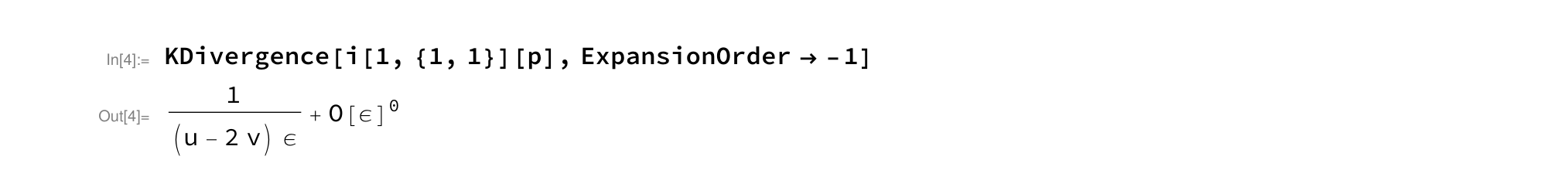}

\noindent drops finite, scheme-dependent terms.

In some cases it may be important only to extract a specific type of singularity as defined in section \ref{sec:divs}. One can use option \verb|Type| to specify types of divergences that should be taken into account. For more examples refer to the notebook \verb|BasicExamples.nb|.

Finally, a double- or triple-$K$ integral is divergent if any of the numbers defined in \eqref{defn} is a non-negative integer. To check if a given integral is divergent use

\includegraphics*[trim=2.0cm 24.5cm 0cm 2.3cm]{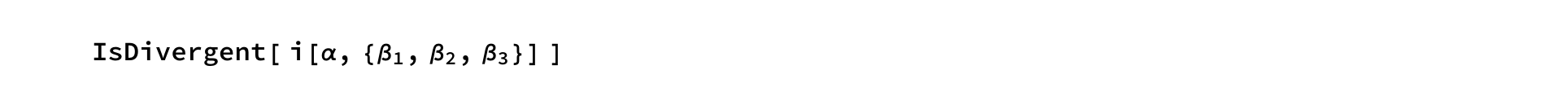}

\noindent which returns \verb|True| if the integral is divergent and \verb|False| otherwise.

\subsection{Momentum loop integrals}

The package represents 1-loop momentum space integrals by \verb|LoopIntegral|. The 2-point and 3-point function loop integrals of the form
\begin{align} \label{23pt}
\int \frac{\D^d \bs{k}}{(2 \pi)^d} \frac{\text{numerator}}{k^{2 \delta_1} |\bs{k} - \bs{q}|^{2 \delta_2}}, \qquad\qquad \int \frac{\D^d \bs{k}}{(2 \pi)^d} \frac{\text{numerator}}{|\bs{k} + \bs{p}_2|^{2 \delta_1} |\bs{k} - \bs{p}_1|^{2 \delta_2} k^{2 \delta_3}}
\end{align}
are represented by

\includegraphics*[trim=2.0cm 24.0cm 0cm 2.3cm]{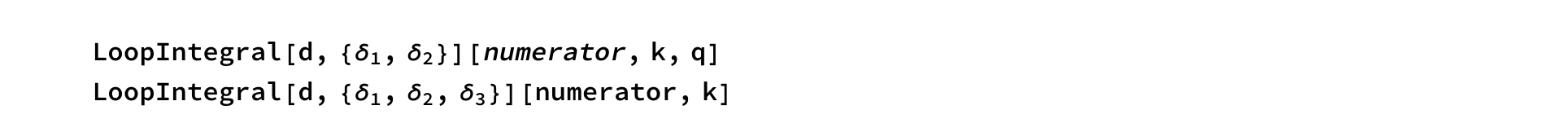}

\noindent respectively. \emph{Note the order of the $\delta$ parameters and a difference in the number of arguments!} In 2-point loop integrals the integral depends on a single external momentum $\bs{q}$, which is specified as a parameter. For the 3-point integrals the external momentum specification is absent: the integral is assumed to depend on $\bs{p}_1, \bs{p}_2$ and $\bs{p}_3 = - \bs{p}_1 - \bs{p}_2$. The integrals can be nested into other integrals providing a framework to write down and evaluate multiple-loop integrals.

In order to express loop integrals in terms of triple-$K$ integrals use

\includegraphics*[trim=2.0cm 24.5cm 0cm 2.3cm]{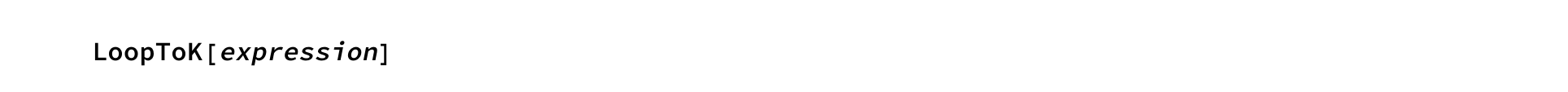}

\noindent By default the function recursively deals with all nested loop integrals and in the process all double-$K$ integrals are reduced to explicit expressions. This may weigh on the performance if the expression actually does not contain any nested integrals. In such a case option \verb|Recursive -> False| can be added, which tells \verb|LoopToK| not to look for any nested integrals. In addition, double-$K$ integrals will not be automatically evaluated.

The result of \verb|LoopToK| can be reduced to the analytic expression by \verb|KEvaluate|. In order to go directly from loop integrals to analytic expressions use

\includegraphics*[trim=2.0cm 24.5cm 0cm 2.3cm]{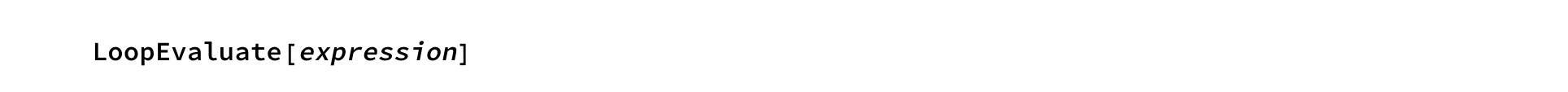}

\subsection{Simplification and manipulations} \label{sec:simp}

Expressions containing loop integrals and multiple-$K$ integrals can be expanded to various extents by two functions

\includegraphics*[trim=2.0cm 24.0cm 0cm 2.3cm]{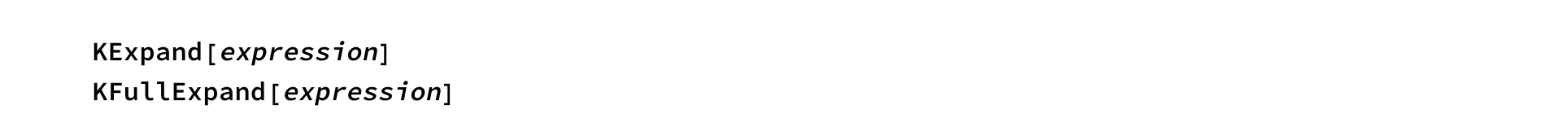}

\noindent By default \verb|KExpand| resolves all derivatives of multiple-$K$ integrals, as well as derivatives of functions \verb|NL| and $\lambda$. It does not substitute analytic expressions for these functions. On the other hand \verb|KFullExpand| fully expands \verb|NL| and $\lambda$ in terms of momenta magnitudes $p_1, p_2, p_3$ according to \eqref{NL} and \eqref{lambda}.

The level of expansion can be controlled by option \verb|Level| to \verb|KExpand|. Level \verb|1| (can also be denoted by \verb|D| or \verb|Diff|) resolves derivatives only and is equivalent to \verb|KExpand| without any options. Levels \verb|2| and \verb|3| (denoted also by \verb|Integer| and $\lambda$) apply various levels of expansion to $\lambda$. Level \verb|4| (also \verb|NL|) fully expands expressions and is equivalent to \verb|KFullExpand|. Level \verb|0| does no expansion.

Basic algebraic relations between various triple-$K$ integrals and loop integrals can lead to significant simplifications. By using

\includegraphics*[trim=2.0cm 24.5cm 0cm 2.3cm]{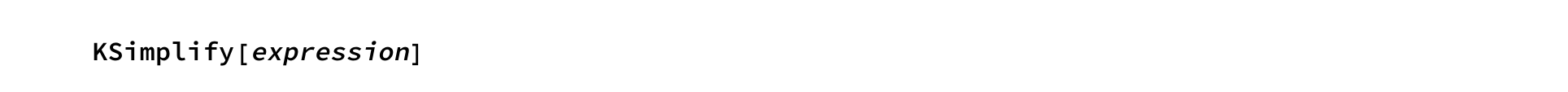}

\noindent the package tries to apply various algebraic relations to carry out such simplifications. Option \verb|Assumptions| can be added in order to supplement \verb|KSimplify| with additional assumptions.

\subsection{Examples}

For a set of neat examples, open the attached notebook \verb|BasicExamples.nb|. Here I will present a rudimentary example of the conformal 3-point function of a single scalar operator $\O$ of dimension $\Delta = 2$ in $d = 3$ spacetime dimensions. In a general CFT such a correlation function in momentum space is given by \eqref{3pt_scalar} up to a multiplicative OPE constant, \verb|C|. Using \verb|KEvaluate| we can compute this correlator with ease:

\includegraphics*[trim=2.0cm 22.8cm 0cm 2.3cm]{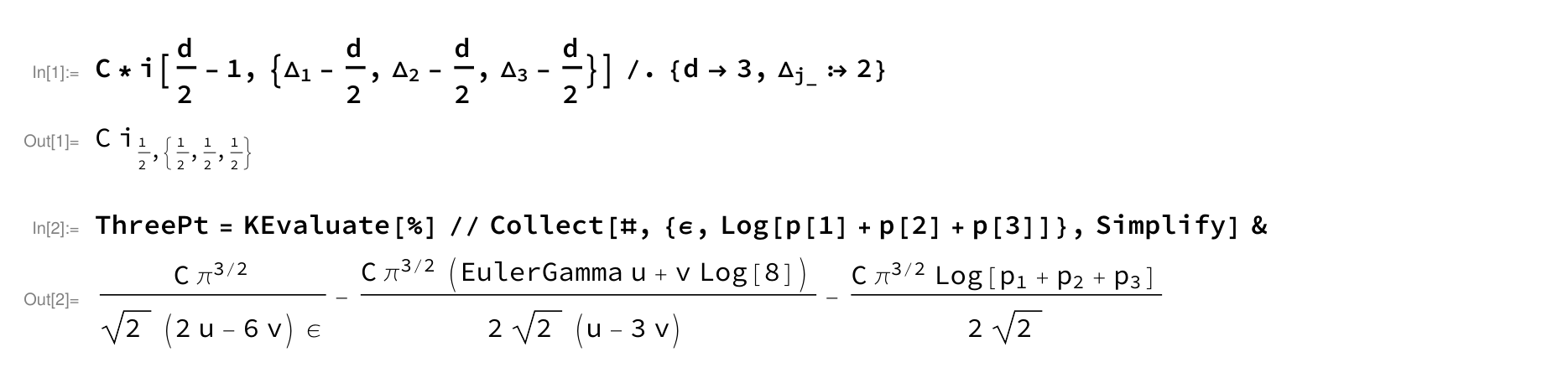}

\noindent Since the triple-$K$ integral $I_{\frac{1}{2} \{ \frac{1}{2} \frac{1}{2} \frac{1}{2} \}}$ representing the correlator diverges, \verb|KEvaluate| used the default regularization \eqref{defreg}. Physically, the divergence indicates the conformal anomaly: the correlator should be rendered finite by the addition of the counterterm of the form $\int \D^d \bs{x} \, \mu^{-(u-v)\epsilon} \phi_0^3$, where $\phi_0$ is the source for the operator $\O$ and $\mu$ the renormalization scale, which must be inserted on dimensional grounds. For the discussion of physics consult example 7 in \cite{Bzowski:2015pba}. Here we will simply use the observation that the term proportional to the logarithm is scheme-independent: it does not depend on neither $u$ nor $v$. Again, the physics picture is that the scaling anomaly is a physical, scheme-independent quantity.

\begin{figure*}[ht]
\includegraphics[width=0.65\textwidth]{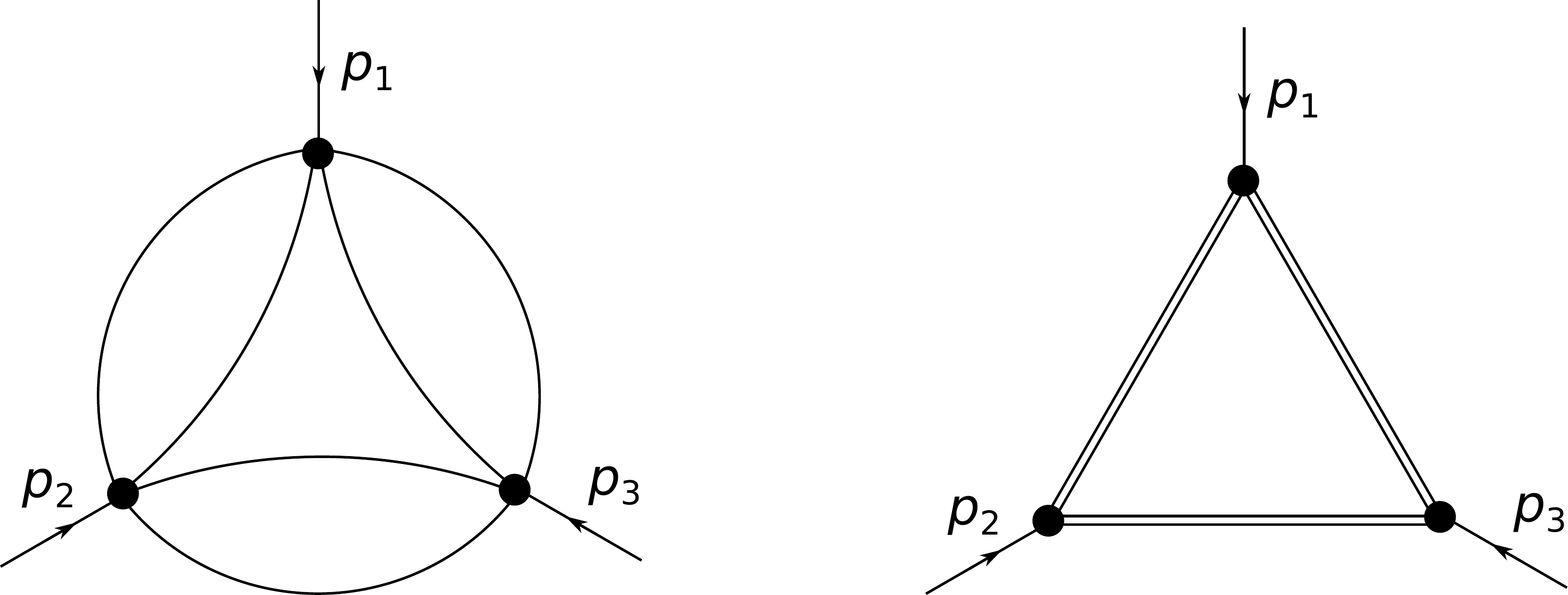}
\centering
\caption{Feynman diagrams representing the 3-point function $\lla \lwick \j^4 \rwick \: \lwick \j^4 \rwick \: \lwick \j^4 \rwick \rra$. Each internal line in the left panel represents the standard massless propagator $1/p^2$. Each double line in the right panel represents the effective propagator obtained from integrating the single loop.\label{fig:1}}
\end{figure*}

Next, we will consider a free massless real scalar field $\j$ and compute the correlation function $\lla \lwick \j^4 \rwick \: \lwick \j^4 \rwick \: \lwick \j^4 \rwick \rra$ in $d=3$ spacetime dimensions. Since the scalar field has dimension $\tfrac{1}{2}$ in $d=3$, the operator $\lwick \j^4 \rwick$ has dimension $2$. Hence, we should recover the expression above and we should be able to calculate the OPE coefficient, \verb|C|, for this particular model.

The Feynman diagram corresponding to the correlation function in momentum space is presented in the left panel of figure \ref{fig:1}. Each 2-point loop can be integrated to yield the effective propagator, denoted by the double line in the right panel. Since we expect that the final 3-point function exhibits singularity, we will work in the standard dimensional regularization scheme with $d = 3 - 2 \epsilon$ and unaltered propagators of $1/p^2$. We find

\includegraphics*[trim=2.0cm 22.8cm 0cm 2.3cm]{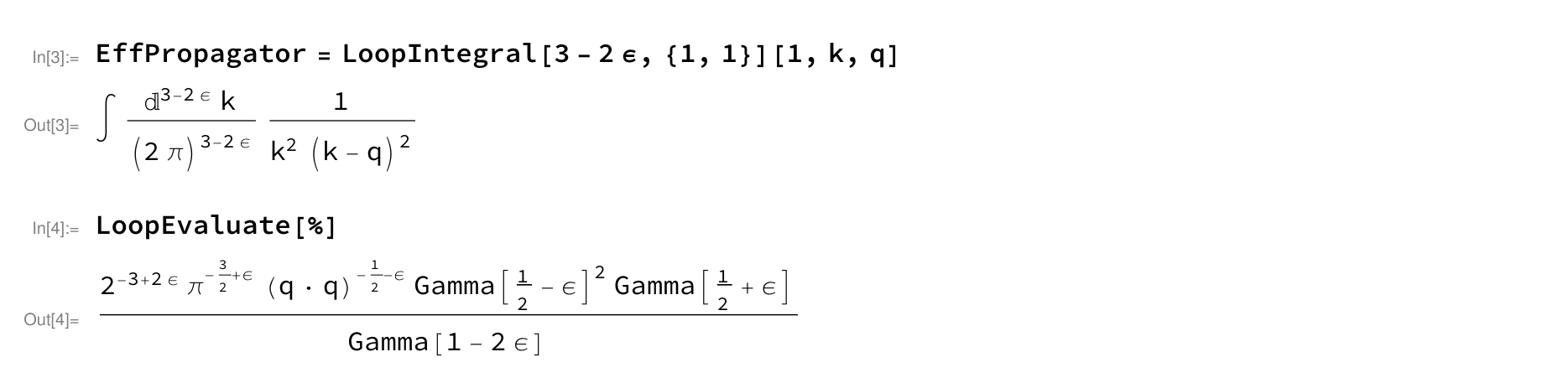}

\noindent The effective propagator is finite in the $\epsilon \rightarrow 0$ limit. However, since we expect the 3-point function to be divergent, we should keep the regulator consistently at each step.

Including the symmetry factor of $[ \binom{4}{2} 2! ]^3 = 1728$ we can write down the loop integral representing the 3-point function and reduce it to the triple-$K$ integral,

\includegraphics*[trim=2.0cm 21.5cm 0cm 2.3cm]{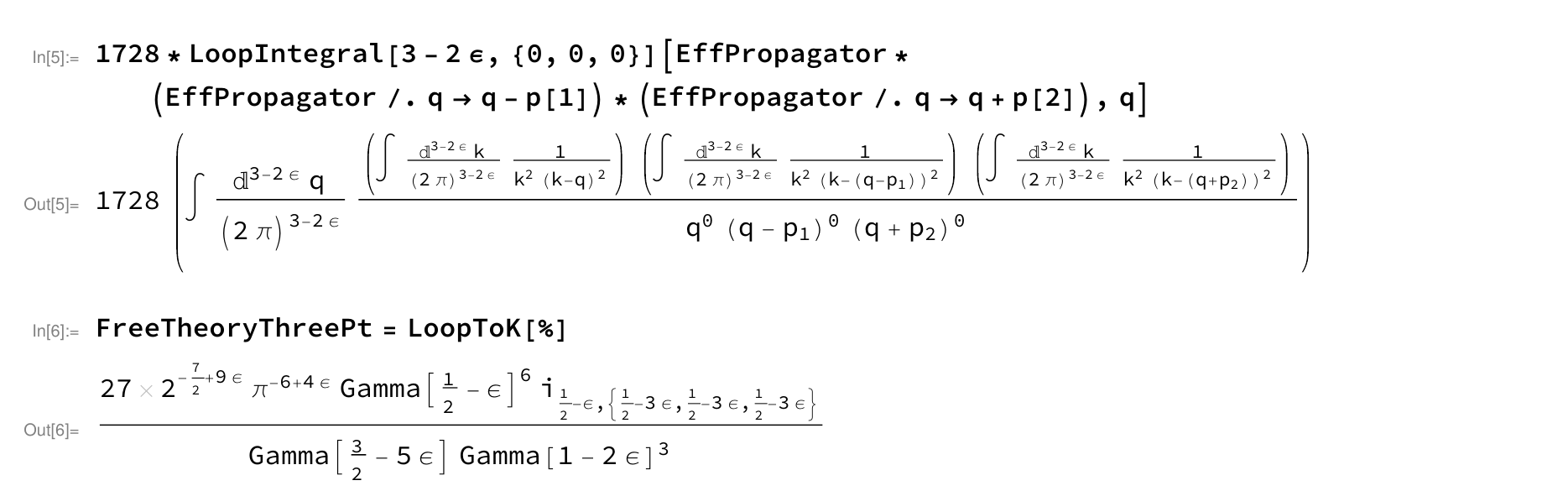}

\noindent In this last expression we recognize the triple-$K$ integral $I_{\frac{1}{2} \{ \frac{1}{2} \frac{1}{2} \frac{1}{2} \}}$ regulated in the specific scheme determined by the use of conventional dimensional regularization.

Finally, we can use \verb|KEvaluate| to evaluate the triple-$K$ integral. As the resulting expression matches the conformal 3-point function stored in \verb|ThreePt|, we can calculate the OPE coefficient by extracting the coefficient of the logarithm.

\includegraphics*[trim=2.0cm 22.5cm 0cm 2.3cm]{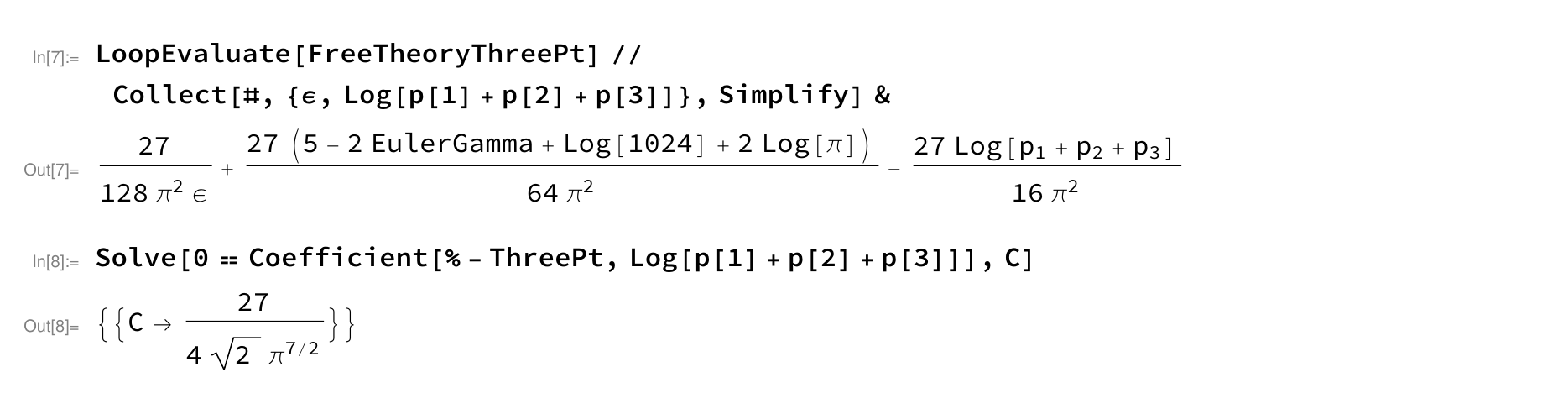}

\section{Konformal.wl}

The package \verb|Konformal.wl| serves as a repository of results regarding conformal 3-point functions of scalar operators, conserved currents and stress tensor. The package contains bulk of the results published in the series of papers, \cite{Bzowski:2013sza,Bzowski:2015pba,Bzowski:2017poo,Bzowski:2018fql}. The package also provides operators present in the analysis of conformal Ward identities.

\subsection{Conformal Ward identities}

The package defines a number of differential operators used in the analysis of conformal invariance. Single scalar $\K_j(\beta)$ operator \eqref{KOp} is represented by \verb|KOp|. Use

\includegraphics*[trim=2.0cm 24.5cm 0cm 2.3cm]{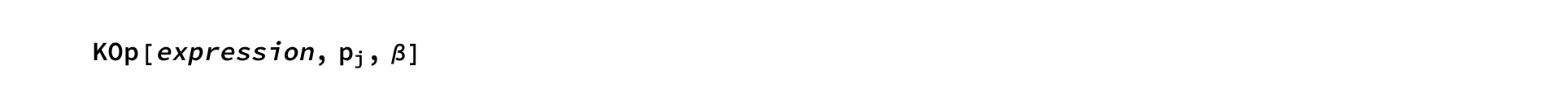}

\noindent to apply $\K_j(\beta)$ to the given expression with parameter $\beta$. Similarly, for the difference $\K_i(\beta_i) - \K_j(\beta_j)$ use

\includegraphics*[trim=2.0cm 24.5cm 0cm 2.3cm]{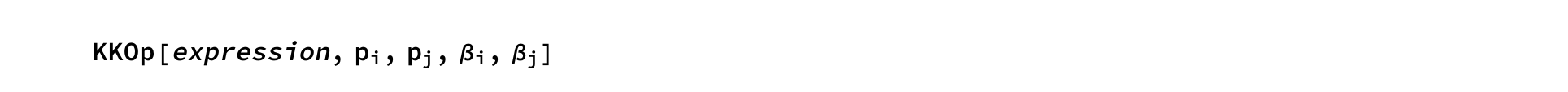}

The full conformal operator as acting on the correlation function in \eqref{CWIOp} can be applied to the given expression by using

\includegraphics*[trim=2.0cm 24.0cm 0cm 2.3cm]{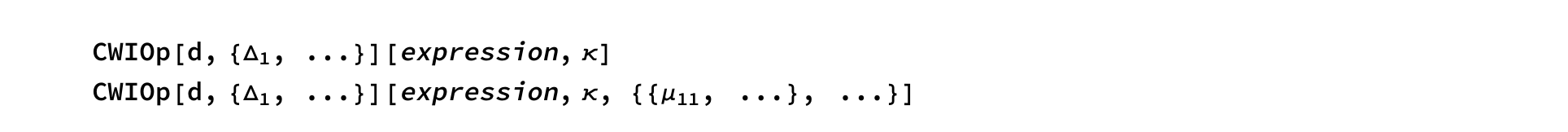}

\noindent In the first form it is assumed that the expression represents an $(n+1)$-function of scalar operators. This means that only the $\mathcal{K}^\kappa$ operator \eqref{CWIOpScal} is applied to the expression. In its second form both $\mathcal{K}^\kappa$ and the spin-dependent part $\mathcal{\bs{T}}^\kappa$ is applied under the assumption that the $j$-th conformal operator has spin $m_j$ as indicated by the list of indices $\mu_{j1}, \ldots, \mu_{j m_j}$. In both cases dimensions of the operators involved are $\Delta_1, \ldots, \Delta_n$ as indicated by the list. The dimension and spin of the last, $(n+1)$-st operator is irrelevant.

In the process of deriving and analyzing conformal Ward identities, a number of differential operators have been introduced. These are:
\begin{itemize}
\item \verb|LOp| and \verb|LprimeOp| denoting operators $L_{s,N}$ and $L'_{s,N}$ as defined in \cite{Bzowski:2013sza}.
\item \verb|ROp| and \verb|RprimeOp| denoting operators $R_{s}$ and $R'_{s}$ as defined in \cite{Bzowski:2013sza}.
\end{itemize}

Finally, solutions to the primary Ward identities stored in \verb|PrimarySolutions| are given in terms of reduced triple-$K$ integrals defined as \eqref{Jint}. In order to convert all $J$-integrals to the standard triple-$K$ integrals in a given expression use

\includegraphics*[trim=2.0cm 24.5cm 0cm 2.3cm]{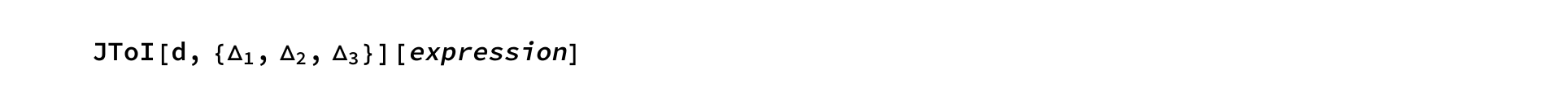}

\subsection{Lists of results}

The following results are stored in \verb|Konformal.wl|.
\begin{itemize}
\item The list of primary Ward identities accessible by \verb|PrimaryCWIs|.
\item The list of secondary Ward identities. Left and right hand sides of the Ward identities can be obtained by \verb|SecondaryCWIsLhs| and \verb|SecondaryCWIsRhs| respectively.
\item The list of transverse Ward identities in generic cases accessible by \verb|TransverseWIs|.
\item The list of solutions to primary Ward identities obtainable by \verb|PrimarySolutions|.
\end{itemize}
In order to access any of the objects, use the corresponding function with the index symbol indicating the 3-point function of interest. The following ten index symbols can be used, corresponding to the obvious correlators:

\includegraphics*[trim=2.0cm 24.5cm 0cm 2.3cm]{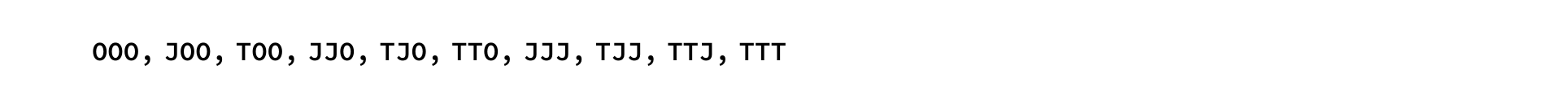}

For example, the list of primary Ward identities for the correlator of a single conserved current and two scalar operators can be obtained by

\includegraphics*[trim=2.0cm 24.0cm 0cm 2.3cm]{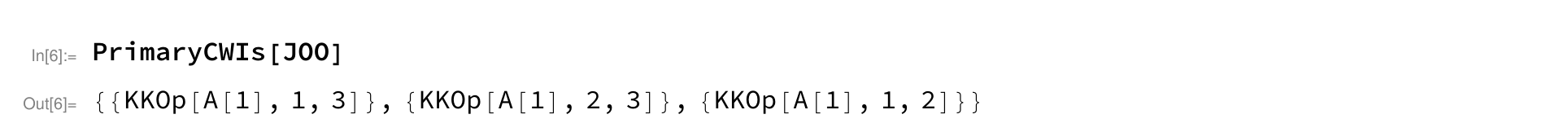}

By default any form factor is denoted by \verb|A|, primary constants by $\alpha$ and any vector index by $\mu$. These can be changed by including options as the arguments to the functions listed above. The options are \verb|FormFactor|, \verb|PrimaryConstant|, and \verb|Index| respectively. For decompositions of tensorial correlation functions and precise definitions of form factors and primary constants, consult \cite{Bzowski:2013sza}.

\subsection{Examples}

The derivation and various checks on all conformal Ward identities can be found in the attached files \verb|DeriveCWIs.nb|, \verb|SolveCWIs.nb| and \verb|CheckCWIs.nb|. Here, just as a quick example, consider the primary solution to the conformal Ward identities in case of the $\< J^\mu \O \O \>$ correlator. The solution can be listed by \verb|PrimarySolutions| and substituted to primary conformal Ward identities stored in \verb|PrimaryCWIs|,

\includegraphics*[trim=2.0cm 22.0cm 0cm 2.3cm]{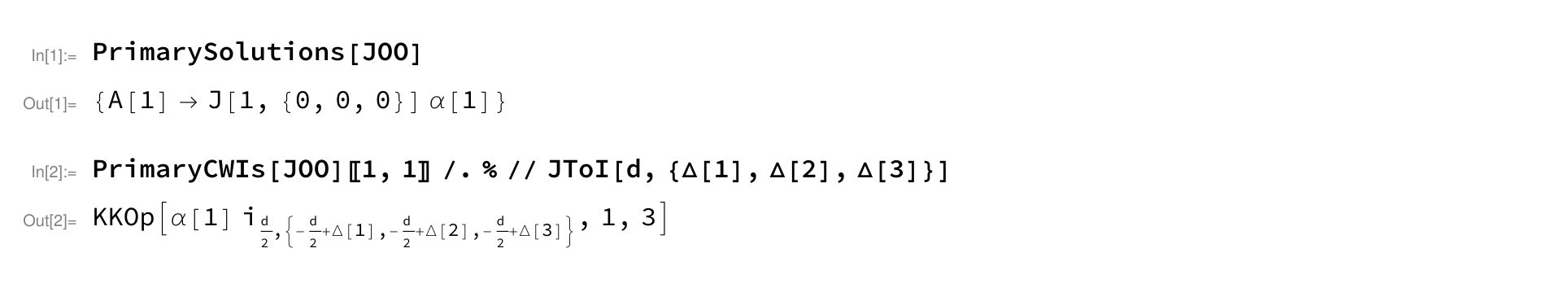}

To evaluate the \verb|KKOp| operator we have to release hold and process the resulting expression. We can use \verb|KExpand| to resolve derivatives of the triple-$K$ integrals, but one must use \verb|KSimplify| to apply suitable identities between various integrals,

\includegraphics*[trim=2.0cm 17.5cm 0cm 2.3cm]{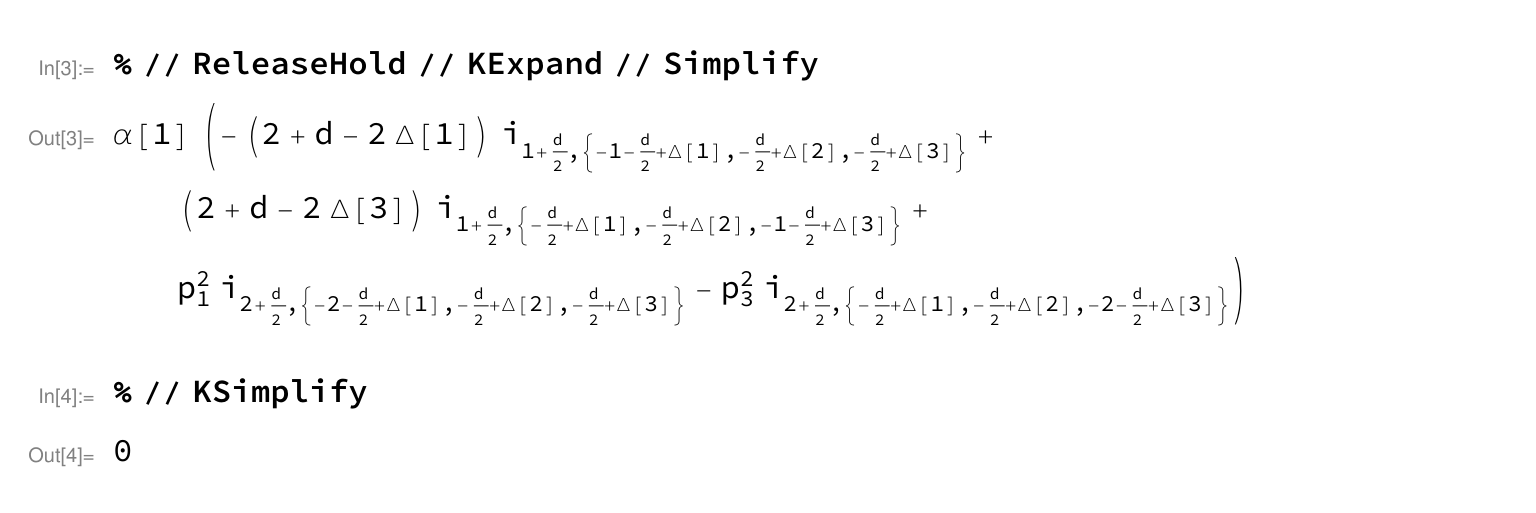}

The expression vanishes indicating conformal invariance of the form factor $A_1$.

\section{Summary}

In this paper I have introduced a Mathematica package designed for manipulations and evaluations of triple-$K$ integrals and conformal correlation functions in momentum space. This includes tools for evaluation of a large number of 2- and 3-point massless multi-loop Feynman integrals with generalized propagators. The package is accompanied by five Mathematica notebooks containing detailed calculations constituting bulk of results published in the sequence of papers \cite{Bzowski:2013sza,Bzowski:2015pba,Bzowski:2017poo,Bzowski:2018fql}.

A number of extensions and features could be added in future. One important direction would be merging the package with functionality provided by well-known packages for loop integral manipulations such as \verb|FeynCalc|, \cite{Mertig:1990an,Shtabovenko:2020gxv} or \verb|Package-X|, \cite{Patel:2015tea,Patel:2016fam}. As far as the content of the package is concerned, extensions to 4-point functions should be possible. Not only would this would include quadruple-$K$ integrals, but also exchange Witten diagrams, \cite{Witten:1998qj}, which represent scattering amplitudes in anti-de Sitter spacetimes. Such results would be beneficial both for investigations in conformal field theory as well as in amplitude-oriented research.

\section*{Acknowledgments}

I am supported by the Knut and Alice Wallenberg Foundation under grant 113410212. I would like to acknowledge years of collaboration with Kostas Skenderis and Paul McFadden in research on triple-$K$ integrals and conformal correlation functions in momentum space.

\end{document}